\documentclass[a4paper,12pt]{article}

\def\ad   {a^{\dagger}}

\def\HP   {\hat {P}}

\def\al   {\alpha}

\def\del  {\delta}

\def\HH {\hat H}

\def\HJ {\hat J}

\def\HP {\hat P}

\def\HH {\hat H}

{\it}
{\it}
{\it}
{\it}
{\it}
{\it}

\def\de {\delta}

\def\Om {\Omega}

\usepackage{times}
\usepackage{graphicx}
\begin{document}
\title{ Hybrid Multideterminant calculation of energy levels of carbon isotopes
   with a chiral effective nucleon-nucleon interaction.}
\author{G. Puddu\\
       Dipartimento di Fisica dell'Universita' di Milano,\\
       Via Celoria 16, I-20133 Milano, Italy}
\maketitle
\begin {abstract}
         We perform  calculations for the binding energies and low-lying
         levels of ${}^{10,11,12,13,14,15,16,17,18,19,20,21,22}C$ nuclei
         starting from the chiral $N3LO$ nucleon-nucleon potential
         within the framework of the Hybrid Multideterminant scheme.
         The effective interaction is obtained using the Lee-Suzuki 
         renormalization scheme applied to 4. and in some case to 5, 
         major harmonic oscillator shells.
         The results are compared with the experimental data.
\par\noindent
{\bf{Pacs numbers}}: 21.60.De, $\;\;\;\;$ 27.20.+n $\;\;\;\;$ 27.30.+t 
\vfill
\eject
\end{abstract}
\section{ Introduction.}
     With the advent of modern accurate nucleon-nucleon interactions and modern 
     many-body computational schemes,  nuclear structure calculations
     starting from the nucleonic degrees of freedom have become possible in the recent years.
     A major advancement has been the systematic construction of realistic nucleon-nucleon
     potential using chiral effective field theories  which start from the most general lagrangian,
     consistent with the symmetries of QCD and the spontaneously broken chiral symmetry,
     appropriate for low energy nucleons and pions (ref.[1]-[4]).
     Using these nucleon-nucleon interactions
     and sometimes even the three nucleon interaction derived from chiral effective field 
     theory several nuclear structure calculations have been performed (ref. [5]-[8]]).
     Typically these calculations are limited to light nuclei ($A\simeq 16$) and in some
     cases, to closed shell medium mass nuclei (ref.[8]). The nuclear structure methods mostly used
     are the no core shell model (ref.[9]-[12]) which pioneered ab-initio nuclear structure calculations, 
     the coupled cluster method (ref.[13]-[15]), the hyperspherical harmonics method (ref.[16][17])
     and, to a lesser extent, the hybrid multideterminant method (HMD) (ref. [18]-[20]). The no core shell 
     model method is limited
     by the size of the Hilbert space which become gigantic as the particle number
     is increased and is  used for $A\simeq 16$. The coupled-cluster method is used
     typically at or around shell closure, but it has been applied also to medium mass
     nuclei (ref. [15]). The hyperspherical harmonics method has been used for very light systems.
     The HMD method, which is utilized in this work, is not limited by 
     the size of the Hilbert space as it can be easily used for medium mass nuclei, 
     and it is equally applicable to closed and open shell nuclei (ref. [21]). Using realistic
     nucleon-nucleon interactions, so far it has been used only in few cases.
     It is our goal to systematically apply this method to nuclei in several mass regions.
     This method belongs to the same family of the VAMPIR methods (ref.[22]-[24]), except that 
     the HMD uses a linear combination of particle Slater determinants instead of
     quasi-particle Slater determinants as in the VAMPIR methods. It is similar
     to the Quantum Monte Carlo method (ref.[25]-[27]), except that the variational method is not 
     stochastic. It utilzes  quasi-newtonian methods (ref.[28]), and the Slater determinants are parametrized 
     differently.
     In this work we take the N3LO nucleon-nucleon interaction (ref.[4]),and study the carbon 
     isotopes, both even and odd, and evaluate ground state energies and few excited states
     for all isotopes under study. Because of the large amount of calculations
     involved, especially for the odd isotopes, we limit ourselves to few  harmonic 
     oscillator frequencies, and renormalize the interaction up to 4, and in some cases up to 5,
     harmonic oscillator
     shells using the Lee-Suzuki (ref.[28]-[31]) renormalization procedure. In an ab-initio 
     approach, one considers several harmonic oscillator frequencies and an increasing  number 
     of harmonic oscillator major shells until the results are independent from the frequency and
     the number of major shells. In practice, at least for this chain of isotopes, this  has never been 
     done so far.
     Such an approach would be necessary if both accurate binding
     energies and excitation energy are required. Here we focus mostly on excitation energies and energy
     differences for which convergence is faster.  
\par
     Some carbon isotopes  have been
     considered in the framework of the UMOA renormalization prescription
     and shell model diagonalization (ref. [32]) with a truncation in the number of allowed excitations.
     More recently they have been  been considered in ref.[33], although the renormalization 
     method is applied in momentum space (with a sharp cutoff at $2.1 fm^{-1}$) rather than in the 
     harmonic oscillator space. Moreover
     an inert ${}^{14}C$ has been assumed  and the neutron single-particle space is 
     restricted to the $sd$ shell. In contrast, we use a no core approach up to the $fp$ shell
     included, and in some case up to $sdg$ shell, with the effective interaction constructed for this 
     space.
     The heaviest of the carbon isotopes ${}^{22}C$ has been recently found to be a 
     borromean nucleus (ref.[34]), that is, stable for particle emission although ${}^{21}C$ is particle
     unstable. 
     All calculations discussed in this work have been performed using personal computers,
     two quad-core and four dual-core processors.
     The outline of this work is the following. In section 2 we give a brief recap
     of the HMD method. In section 3 and subsections we discuss the results and compare
     with the experimental data. In section 4 we summarize the results.
\bigskip
\bigskip
\section{ A brief recap of the HMD method.}
\bigskip
\bigskip
     The HMD method (ref.[18]-[20]) consists in solving the many-body Schrodinger 
     equation using as ansazt for the yrast eigenstates $|\psi>$ a linear combination of Slater 
     determinants, i.e.
$$
|\psi>=\sum_{\al=1}^{N_w} g_{\al}\HP |\phi,\al>.
\eqno(1)
$$
     The operator $\HP$ restores the desired exact quantum numbers (angular momentum and 
     parity), $\al$ labels the Slater determinants and $|\phi,\al>$ is a general Slater determinant 
     ( that is, no symmetries are imposed).
     Each Slater determinant is built from the generalized creation operators
$$
c^{\dagger}_n(\al)=\sum_{i=1}^{N_s} U_{i,n}(\al)\ad_i.
\eqno(2)
$$
     $\ad_i$ is the creation operator in the harmonic oscillator single-particle state $i$, and
     $N_s$ is the dimension of the single-particle basis. The complex numbers $g_{\al}$ and
     $U_{i,n}$ are determined by minimizing the expectation values of the Hamiltonian with
     quasi-newtonian methods ( cf. ref. [28],[35] and references in there). Clearly the larger the number of 
     Slater determinants $N_w$ the more
     $|\psi>$ will approach the exact yrast eigenstate. The ansatz of eq.(1) is valid for
     yrast eigenstates, for excited eigenstates having the same quantum numbers we must in addition
     add terms containing the lower eigenstates with the same quantum numbers and the linear
     combination must preserve orthogonality with the previously determined eigenstates (ref.[23]).
\par
     The degree of accuracy of the ansatz of eq.(1) for finite $N_w$ has been recently analized in 
     ref. [36] in order to construct extrapolation techniques, using the phenomenological 
     $fpd6$ realistic effective interaction. We have tested the
     accuracy and effectiveness of our quasi-newtonian variational method for ${}^{56}Ni$. 
     Using $15$ angular momentum projected Slater determinants we obtained for the ground state energy 
     $-203.157 MeV$ and with  $25$ Slater determinants we obtained $-203.175 MeV$ and using $35$
     Slater determinants $-203.182 MeV$. This is
     to be compared  with the exact shell model value of $-203.198 MeV$ quoted in ref.[36],
     and with the Quantum Monte Carlo result (prior extrapolation) of $-203.161 MeV$
     which was obtained with  $150$ Slater determinants (ref. [36]).
\par
     In practice, for ab-initio no-core calculations, we avoid the use of the full angular momentum 
     projector since 
     experience shows that, in such cases, 
     we need a rather large number of fully angular momentum and parity projected Slater determinants
     to obtain good approximations to the eigenstates and therefore, in order to reduce the 
     computational cost, we proceed as follows.
\par\noindent
     We add to the
     Hamiltonian a term $\gamma( \hat J^2-J(J+1))$ where $\HJ$ is the angular momentum operator
     and $J$ is the desired value 
     and we use, instead of the full angular momentum projector only the projector to good
     projection onto the z-axis $J_z=J$, much in the same way it is done in standard shell model
     calculations. This device is very useful especially for odd and odd-odd mass nuclei.
     The wave functions obtained in this way are used to evaluate observables with the full
     three-dimentional angular momentum projector.
     Experience shows
     that few hundreds Slater determinants are relatively easy to obtain and the full re-projection of the 
     wave function
     obtained this way is much less expensive than the use of the full projector from the beginning.
     However, if we desire excited states with the same exact quantum numbers, the use of the full
     projector  seems necessary so far.
\par 
     As discussed in the next section, for no-core calculations, we need several hundreds $J_z^{\pi}$ 
     projected Slater determinants
     to reach a reasonable convergence to the energies, however the convergence to the excitation
     energies is much faster, provided wave functions with different $J_z^{\pi}$ undergo exactly
     the same sequence of computational steps. The number of Slater determinants necessary to achieve
     convergence increases with the  number of major shells. Hence for 5 major shells calculations
     we only evaluate excitations energies. 
\par
     The intrinsic Hamiltonian used in the calculations is obtained in the following way.
     First an harmonic oscillator potential is added to the A-particle Hamiltonian,
     the resulting Hamiltonian is A-dependent. The two-body interaction is obtained by
     renormalizing the two-particle A-dependent Hamiltonian with the Lee-Suzuki procedure, much
     in the same way it is done in the no core shell model (cf. ref. [9] for a detailed description).
     The two-particle interaction is  restricted to some number of relative coordinate 
     harmonic oscillator shells $N_r+1$. The two-body matrix elements of the intrinsic Hamiltonian for the 
     A-particle system can then be constructed. Using the Talmi-Moshinski
     transformations brackets, the matrix elements of this intrinsic Hamiltonian are evaluated
     up to $N_r/2+1$ major shells in the frame of the single-particle coordinates. This is the HMD-a version of the method
     (cf. ref. [20] for more details). Usually, in order to prevent center of mass excitations
     in the evaluation of excited states a term proportional to the harmonic oscillator
     Hamiltonian for the center of mass
     $\beta(\HH_{cm}-3/2\hbar\Om)$ is added at the end.
\par 
     The variational calculation is carried out progressively. That is, we start with a single
     Slater determinant and add trial Slater determinant one at a time and always optimize the
     last added Slater determinant.
     At specific numbers of Slater determinants we vary anew all Slater determinants one at
     a time. For example when $N_w=5$
     all Slater determinants are varied anew, and after we reach, say, $N_w=10,15,25,35,50 $ we re-optimize
     all Slater determinants etc.. The numbers 
     $5,10,15,25,35,70,100,150...$ are somewhat a free choice. By far this is the most
     expensive part of the calculation, especially if we consider the full angular momentum and
     parity projector, hence the choice to replace it with a partial projector and only after
     a sufficiently large number of Slater determinants has been constructed, we use the full projector to evaluate 
     expectation values.
\par
     Both the method and the set of computer codes have been extensively tested.
\section{ Carbon isotopes.}
\par 
    For all the cases discussed below the coefficient  of the center of mass Hamiltonian
    is fixed to $\beta=0.7$, the harmonic oscillator frequency is for most of the cases  
    $\hbar\Om=14 MeV$.
    The coefficient $\gamma$ of the $\hat J^2-J(J+1)$ term is set to $0$ for the even-even
    isotopes to $2 MeV$ or $4 MeV$ for the odd-mass isotopes. 
    In the following, the experimental values of the binding energies are taken
    from ref. [37] and the excitation energies form ref. [38]. For ${}^{17}C$ and ${}^{19}C$ 
    the experimental data is taken from ref. [39],  for ${}^{18}C$ from ref. [40] and for ${}^{20}C$
    from ref. [41].
    In what follows we also discuss the variation of the number of nucleons in the single
    particle shells and define $\del n(E_x,a,t)=n(E_x,a,t)-n(E_{gs},a,t)$ where $E_x$ is any
    of the excited states, $E_{gs}$ is the ground state energy, $a$ is any of the single-particle shells 
    and $t=n,p$ 
    denotes the type of particles (neutrons or protons).
    Only the largest variations will be given, the ones
    that are omitted are too small compared with the others. This is a very simple way 
    to classify the type of excitation, e.g. neutron excitation, proton excitation or both.
\par\noindent
    As previously mentioned, in some cases we have performed calculations also with
    5 major shells. We find that the absolute binding energies are different from the
    ones obtained with 4 major shells, however the excitation energies are rather similar.
    This reflects the fact that energy differences converge much better than the energies.
    Also the value of $\hbar\Om=14 MeV$ is close to the energy minimum as a function
    of $\hbar\Om$, thereby decreasing the dependence of the energies on  $\hbar\Om$. A
    systematic calculation for several values of $\hbar\Om$, for 4 and 5 major shells,
    for all these isotopes is too lengthy on personal computers. Unless explicitely 
    stated we consider $\hbar\Om=14 MeV$ and 4 major shells.
\par
\subsection{ ${}^{10}C$}
\par
    The experimental binding energy of ${}^{10}C$ is  $60.320 MeV$. Using $250$
    Slater determinants, optimized as explained in the previous section with $J_z^{\pi}=0^+$ and then
    re-projecting to $J^{\pi}=0^+$, in order to evaluate the expectation values, we
    obtained a binding energy of $53.438 MeV$. The behavior of the ground state energy
    as a function of the inverse of the number of Slater determinants, $N_{SD}$, is shown in fig. 1. The 
    behavior
    of the energies in fig. 1 is typical if the number of Slater determinants is large enough. It is reasonable,
    due to the linear behavior of the energy as a function of $1/N_{SD}$ to extrapolate in order
    to estimate the uncertainty of the calculation. The extrapolated value for the ground-state
    energy is $-53.808 MeV$, hence our result has an uncertainty of $0.7\%$. We found this $1/N_{SD}$
    behavior in most of the cases. Only in few cases the number of Slater determinants was not sufficiently
    large.  We  use anyway  a linear extrapolation in order to have an estimate of the uncertainty of 
    the calculations. These uncertainties should not be confused with the statistical uncertainties 
    as in Monte Carlo calculations. They are simply an estimate of the possible decrease of the energies
    if we would increase the number of Slater determinants, that is, how far we are from the exact values.
\par 
\renewcommand{\baselinestretch}{1}
\begin{figure}
\centering
\includegraphics[width=10.0cm,height=10.0cm,angle=-90]{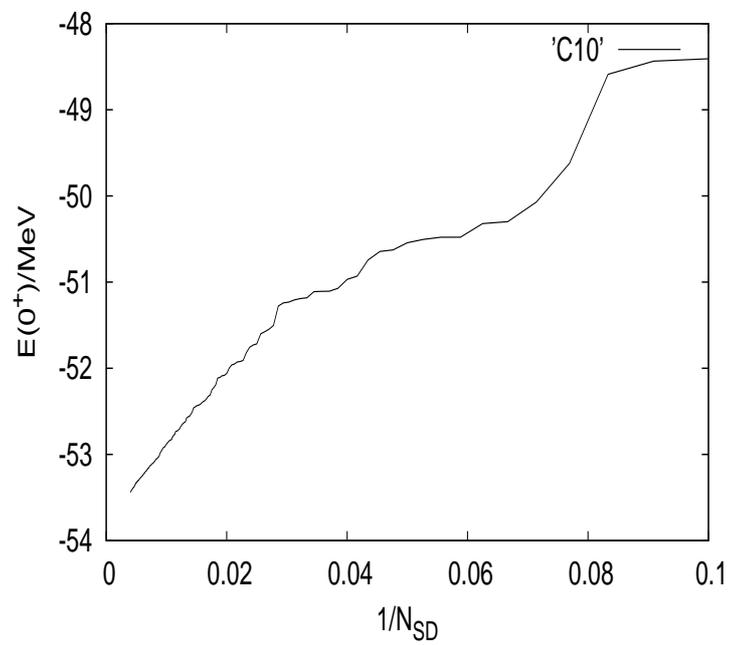}
\caption{Ground-state energy of ${}^{10}C$ as a function of the inverse of the number of Slater determinants.}
\end{figure}
\renewcommand{\baselinestretch}{2}
    For the light carbon isotopes we find that theoretical binding energies are underestimated
    compared to the experimental values, while for the heavy carbon isotopes the theoretical values
    overestimate the corresponding experimental values.
\renewcommand{\baselinestretch}{1}
\begin{figure}
\centering
\includegraphics[width=10.0cm,height=10.0cm,angle=-90]{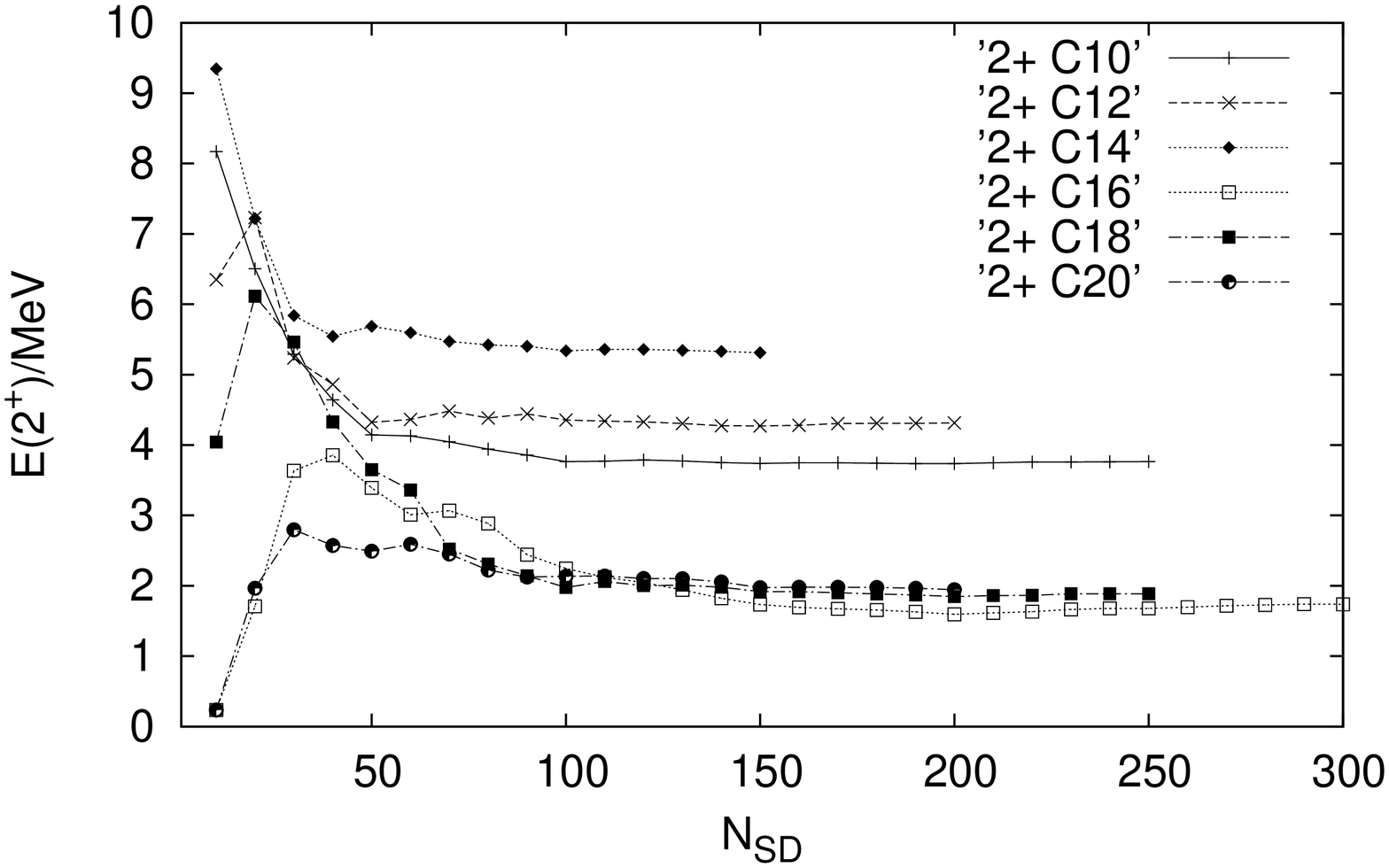}
\caption{Excitation energy of the $2^+_1$ state as a function of the number of Slater 
determinants for the even carbon isotopes.}
\end{figure}
\renewcommand{\baselinestretch}{2}
    The experimental
    value of the excitation energy of the $2^+_1$ state is $3.354 MeV$. Our calculation gives
    $E(2^+_1)=3.764 MeV$. In fig.2 we show the behavior of the excitation energy as a function of
    the number of the Slater determinants. As it can be seen the value of the excitation energy
    is rather stable and the oscillation for large $N_{SD}$ have an amplitude of about $10 keV$
    for this nucleus.
    The reason for this remarkable stability is that both calculations for the $0^+$ and the $2^+$ energy
    have almost the same error and of the same sign (these are variational calculations)
    and this error cancels out in the evaluation of the excitation energy. This is the
    reason why excitation energies converge much better than the absolute value of the energies.
    It should be stressed however that the relative uncertainty in the binding energy is small.
    It is interesting to look at the variation of the population of nucleons as we go from the ground-state
    to the excited state.
    We have $\del n(2^+,0p,n)=0.08$ and $\del n(2^+,0p,p)=0.07$.
    Although small, the number of neutrons (protons) excited above the $s$ and $p$ shells is non-zero:
    $\del n(2^+,sd,n)=0.17$, $\del n(2^+,sd,p)=0.21$ and  $\del n(2^+,fp,n)=0.12$ and 
    $\del n(2^+,fp,p)=0.15$.
\par
    For this nucleus, we performed also a calculation with $\hbar\Om=11 MeV$  with 4 major shells.
    The binding energy becomes $51.372 MeV$ (compared with $53.438 MeV$ for $\hbar\Om=14 MeV$), however
    the excitation energy of the $2^+_1$ is $E(2^+_1)= 3.72 MeV$ and it is well converged as a function
    of the number of Slater determinants, and it is almost the same as the one obtained
    for $\hbar\Om=14 MeV$, which is $3.764 MeV$.
    For $\hbar\Om=17 MeV$,  the excitation energy of the $2^+_1$ state obtained with $300$ Slater
    determinants, becomes 
    $E(2^+_1)= 3.73 MeV$. It is quite remarkable that although the energies have a non negligible
    $\hbar\Om$ dependence, the excitation energies are nearly constant.
\par\noindent
    Using $\hbar\Om=14 MeV$ we have also performed
    a calculation with 5 major shells. However we used only $200$ Slater determinants, and obtained
    $E(2^+_1)=3.67 MeV$. The calculation is not entirely converged since  the excitation energy has
    a small increase with the number of Slater determinants (about $70 KeV$ in the last $30$ Slater 
    determinants), but it is consistent with and it approaches the values obtained with 4 major 
    shells.
    This shows that working with 4 major shells and $\hbar\Om=14 MeV$, gives reliable results for
    the excitation energies for this nucleus.
\par
    At this point a few comments are in order about the convergence of our method. The HMD method
    is applicable regardless of the dimensionality of the Hilbert space, however we do not know
    yet how many Slater determinants we have to optimize in order to obtain the energies within, say $1\%$ 
    accuracy. We do know however that larger Hilbert spaces require a larger number of Slater determinants.
    Using $4$ major shells, we need a few hundreds Slater determinants (perhaps even $500$), for 
    calculations utilizing $5$ major shells this number is higher, hence it is not so surprising
    that the excitation energy for the $2^+_1$ state in the case of $5$ major shells is not entirely
    converged with $200$ Slater determinants. Presumably, the optimal way to calculate binding energies
    is to evaluate differences of binding energies and to perform an accurate binding energy calculation
    on just one isotope. An other possibility is to explore, in the context of ab-initio calculations,
    the extrapolation method of ref.[36].
\par
\subsection{ ${}^{11}C$}
\par
     The experimental value of the binding energy of ${}^{11}C$ is $73.44 MeV$ and the 
     ground state has $J^{\pi}=3/2^-$, which is reproduced by our calculation. The theoretical
     value is $67.842MeV$. As before for large $N_{SD}$ the energy is linear as a function of $1/N_{SD}$.
     and the extrapolated value is $68.546 MeV$ giving a theoretical  uncertainty of $1\%$.
\renewcommand{\baselinestretch}{1}
\begin{figure}
\centering
\includegraphics[width=10.0cm,height=10.0cm,angle=-90]{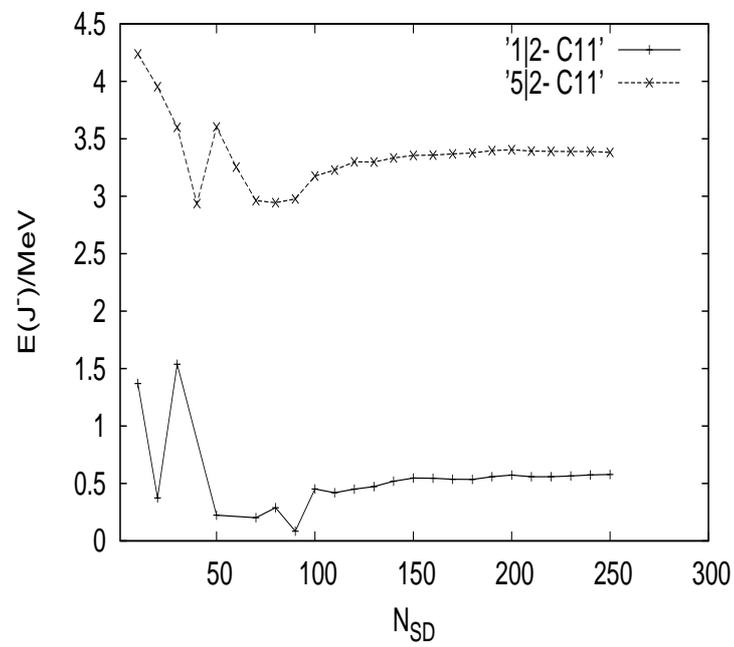}
\caption{Excitation energy of the $1/2^-$ and $5/2^-$ states as a function of the number of Slater 
determinants for ${}^{11}C$.}
\end{figure}
\renewcommand{\baselinestretch}{2}
     The energy of the first excited state ($1/2^-$) is not well reproduced. The experimental 
     value is $2 MeV$, while our calculation gives $0.58 MeV$. The first $5/2^-$ state has
     an experimental excitation energy of $4.32 MeV$, our calculation gives $3.38 MeV$.
     In fig.3 we show the behavior of the excitation energies as a function of the number of
     Slater determinants.
     The number of neutrons (protons) for the ground-state in the $s$, $p$, $sd$ and $fp$ shells are
     $1.81 (1.83), 2.79 (3.75), 0.21 (0.22), 0.19 (0.2)$ respectively. Moreover for the $1/2^-$ state
$$
\del n(1/2^-,0p3/2,n)=-0.19,\;\;\; \del n(1/2^-,0p1/2,n)=0.19
$$
$$
\del n(1/2^-,0p3/2,p)=-0.25,\;\;\; \del n(1/2^-,0p1/2,p)=0.24
$$
     The $5/2^-$ state is primarily a neutron excitation. In fact
$$
\del n(5/2^-,0p3/2,n)=-0.38,\;\;\; \del n(5/2^-,0p1/2,n)=0.38 \\
$$
$$
  \del n(5/2^-,0p3/2,p)=-0.1,\;\;\; \del n(5/2^-,0p1/2,p)=0.09.
$$
\par
\subsection{ ${}^{12}C$}
\par
     ${}^{12}C$ has been extensively investigated, both experimentally and theoretically
     because of its astrophysical importance. As in the no core shell model calculations
     the $0^+_2$ state (the Hoyle state), is missing at low energy. A small number of harmonic
     oscillator major shells is not sufficient to reproduce the position of this state.
     The experimental binding energy is $92.16 MeV$, the calculated value with $200$ Slater determinants
     is $90.154 MeV$ and the extrapolated value is $90.773 MeV$. We performed another calculation
     using $400$ Slater determinants and obtained $90.503MeV$ and a corresponding extrapolated
     value of $90.940 MeV$. In this case the $1/N_{SD}$ behavior seen in the previous cases
     is not entirely correct. This example shows that the extrapolated values give simply
     an uncertainty  of the calculated ones. Also in this case the uncertainty is about $1\%$.
     The calculated excitation energy of the $2^+_1$ state is $4.31 MeV$ to be compared with the 
     experimental value of $4.44 MeV$. The behavior of the excitation energy as a function of the
     number of Slater determinants is show in fig. 2. The occupation numbers for the ground state
     are nearly equal for neutrons and protons, and a small number of neutrons and protons
     is moved from the $0p3/2$ to the $0p1/2$ shell ($<0.1$) for the $2^+$ state.
\par
\subsection{ ${}^{13}C$}
\par
     Odd-mass isotopes allow to study whether the single-particle properties of the Hamiltonian
     are correct. The experimental binding energy of ${}^{13}C$ is $97.11 MeV$ and the ground-state
     has $J^{\pi}=1/2^-$. Some low-lying yrast levels of negative parity are (in MeV)
     $E(3/2^-)=3.68 $ ,$E(5/2^-)=7.55$. 
     Our calculation reproduces	the correct $J^{\pi}=1/2^-$ of the ground-state with a
     binding energy of $97.58 MeV$  (with $250$	Slater determinants and	an uncertainty of $0.5\%$).
     For the above negative parity levels we obtained $E(3/2^-)=2.6 MeV$ and $E(5/2^-)=5.89 MeV$
     Regarding the nature of these states, we have
$$
\del n(3/2^-,0p3/2,n)=-0.34,\;\;\;\del n(3/2^-,0p1/2,n)=0.35
$$
$$
\del n(3/2^-,0p3/2,p)=-0.33,\;\;\;\del n(3/2^-,0p1/2,p)=0.33
$$
     and for the $5/2^-$ state
$$
\del n(5/2^-,0p3/2,n)=0.14,\;\;\;\del n(5/2^-,0p1/2,n)=-0.16
$$
$$
\del n(5/2^-,0p3/2,p)=-0.46,\;\;\;\del n(5/2^-,0p1/2,p)=0.47
$$
     These variations show that the $5/2^-$ is primarily a proton excitation.
\par
\renewcommand{\baselinestretch}{1}
\begin{figure}
\centering
\includegraphics[width=10.0cm,height=10.0cm,angle=-90]{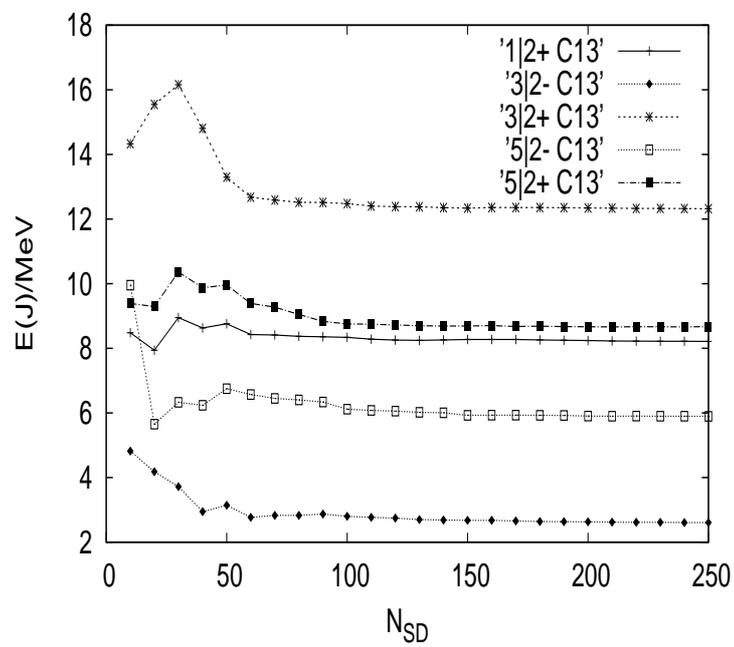}
\caption{Excitation energies of the selected states as a function of the number of Slater 
determinants for ${}^{13}C$.}
\end{figure}
\renewcommand{\baselinestretch}{2}
     The positive parity levels involve the $sd$ shell. The experimental locations in MeV are
     (we consider only few yrast levels) $E(1/2^+)=3.09$, $E(5/2^+)=3.85$ and $E(3/2^+)=7.69$.
     The corresponding theoretical values are $E(1/2^+)=8.2$ , $E(5/2^+)=8.67$ and $E(3/2^+)=12.32$,
     nearly $5 MeV$ too high.
     The variations of the occupation numbers reveal the nature of these levels. We have
$$
\del n(1/2^+,0p3/2,n)=-0.53 ,\;\;\del n(1/2^+,0p1/2,n)=-0.35,\;\;\del n(1/2^+,1s1/2,n)=0.83
$$
$$
\del n(1/2^+,0p3/2,p)=-0.25 ,\;\;\;\del n(1/2^+,0p1/2,p)=0.30
$$
     The neutron $1s1/2$ orbital contains one extra neutron. All others $\de n$  are small.
     For the $5/2^+$ state we have
$$
\del n(5/2^+,0p3/2,n)=-0.53 ,\;\;\del n(5/2^+,0p1/2,n)=-0.34,\;\;\del n(5/2^+,0d5/2,n)=0.94
$$
$$
\del n(5/2^+,0p3/2,p)=-0.29 ,\;\;\;\del n(5/2^+,0p1/2,p)=0.34
$$
     Almost one extra neutron in the $0d5/2$ orbital. For the $3/2^+$ state we have
$$
\del n(3/2^+,0p3/2,n)=-0.59 ,\;\;\del n(3/2^+,0p1/2,n)=-0.28,\;\;\del n(3/2^+,0d3/2,n)=0.78
$$
$$
\del n(3/2^+,0p3/2,p)=-0.34 ,\;\;\;\del n(3/2^+,0p1/2,p)=0.38
$$
     Almost one extra neutron in the $0d3/2$ orbital. In all cases there is a strong 
     proton excitation.
     For this nucleus different value of $\hbar\Om$ were not considered. The rather large
     excitation energy across major shells remains to be understood, that is, whether it is an 
     artifact of the restriction to $4$ major shells, or it is a feature of this $NN$ interaction.
     Eventually, this nucleus
     will be studied in the future in a more detailed way (i.e. a larger number of major shells
     and several values of $\hbar\Om$).
\par
\subsection{ ${}^{14}C$}
\par
     The experimental binding energy of this nucleus is $105.284 MeV$ and the excitation energy
     of the $2^+_1$ state is $7.01 MeV$. The first excited state is a $1^-$ state at $6.09 MeV$.
     This high excitation energy is considered as a motivation for model assumptions that take 
      ${}^{14}C$ as an inert core.
     We considered $150$ Slater determinants. Our result for the binding energy is $109.976$
     with an uncertainty of $0.37\%$. As in the previous cases, the energy shows a $1/N_{SD}$ behavior
     for large $N_{SD}$. Our values for the excitation energies are
     $E(2^+_1)=5.31 MeV$ and $E(1^-)=12.3 MeV$. As expected the $2^+$ state is a proton excitation
     and $\del n(2^+,0p3/2,p)=-0.68$ and $\del n(2^+,0p1/2,p)=0.69$. The number of neutrons
     in the $0s$ and $0p$ shells is $7.3$ indicating that the closure of the neutron shell is partially
     broken. One can see this more explicitely by comparing the final ground-state results with
     the ones obtained with a Hartree-Fock calculation using the full angular momentum projector.
     The  HF  binding energy is $106.33 MeV$, close to the HMD value. However
      the proton occupation numbers for the $p$ shell are different and, to a less extent, also the neutron
     occupation numbers.
     The calculated $1^-$ state is mostly a neutron excitation, in fact
$$
\del n(1^-,0p3/2,n)=-0.17 ,\;\;\del n(1^-,0p1/2,n)=-0.67,\;\;\del n(1^-,1s1/2,n)=0.82
$$
$$
\del n(1^-,0p3/2,p)=-0.21 ,\;\;\;\del n(1^-,0p1/2,p)=-0.25
\eqno(14)
$$
     Both the structure of this state and the high energy of the $1^-$ state again indicate
     that  the distance between the $p$ and $sd$ major shells is too large.
\par
     For this nucleus we performed also a calculation of the excitation energy of the $2^+_1$
     state using $\hbar\Om=11 MeV$ and $200$ Slater determinants. We obtained $E(2^+_1)=4.36 MeV$.
     Using $\hbar\Om=17 MeV$ with $200$ Slater determinants we obtained $E(2^+_1)=5.6 MeV$ (we did not
     in this case reevaluate the excitation energy with the full angular momentum projector).
     These results show a dependence of $E(2^+_1)$ on $\hbar\Om$. We therefore performed 
     for this nucleus a calculation using $5$ major shells. Again no reprojection was performed
     at the end of the calculation for this case. For $\hbar\Om=11 MeV,\; 14 MeV,\; 17 MeV$ we obtained 
     $E(2^+_1)=4.23 MeV,\; 4.8 MeV,\; 4.7 MeV$, respectively. The uncertainty of the calculation
     is about $0.1 MeV$. There is still a residual $\hbar\Om$ dependence of the excitation energy,
     but it is smaller than the one obtained with $4$ major shells.

\subsection{ ${}^{15}C$}
\par
     The experimental binding energy of ${}^{15}C$ is $106.5 MeV$ and the ground state 
     has $J^{\pi}=1/2^+$. The first excited state has $E(5/2^+)=0.74 MeV$ and the first
     $3/2^+$ state is at $4.78 MeV$. The first few negative parity levels are
     $E(1/2^-)=3.10 MeV$, $E(5/2^-)=4.22 MeV$ and $E(3/2^-)=4.66 MeV$. 
     Our results are the following. The binding energy obtained with $250$ Slater determinants
     is $110.586 MeV$ with an uncertainty of $0.46\%$. The ground-state spin and parity are
     properly reproduced. Heavy carbon isotopes overbind compared to the experimental data
     while the light ones underbind.
\renewcommand{\baselinestretch}{1}
\begin{figure}
\centering
\includegraphics[width=10.0cm,height=10.0cm,angle=-90]{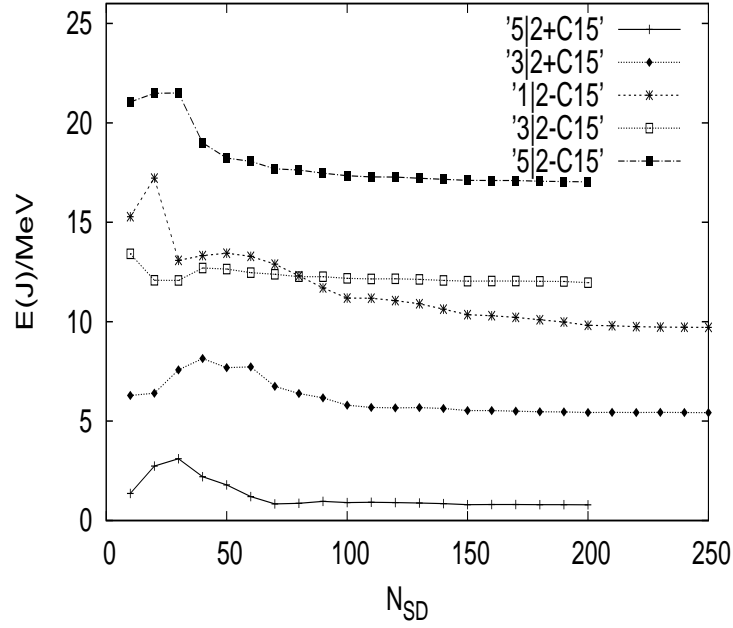}
\caption{Excitation energies of the selected states as a function of the number of Slater 
determinants for ${}^{15}C$.}
\end{figure}
\renewcommand{\baselinestretch}{2}
     Our results for the yrast positive parity levels are: $E(5/2^+)=0.79 MeV$ and 
     $E(3/2^+)=5.43 MeV$. If we compare the occupation numbers of the $1/2^+$ state of ${}^{15}C$
     with the occupation numbers of the ground-state of ${}^{14}C$ we find that mostly
     they differ because of the population of the $1s1/2$ neutron shell. The difference
     in the number of neutrons for this shell is $0.88$. The remaining $0.12$ neutrons
     are accounted for small difference in the population of the other neutron shells.
     The largest differences in the occupation numbers of the $5/2^+$ state and the $1/2^+$ state
     are the following
$$
\del n(5/2^+,0d5/2,n)=0.91 ,\;\;\;\del n(1/2^+,1s1/2,n)=-0.88
$$
$$
\del n(5/2^+,0p3/2,p)=-0.13 ,\;\;\;\del n(1/2^+,0p1/2,p)=0.13
$$
     That is, the $5/2^+$ state is predominantly, but not entirely a neutron excitation.
     The $3/2^+$ state is not a neutron excitation built on the ground state. 
     In fact the dominant differences in the occupation numbers are
$$
\del n(3/2^+,0d3/2,n)=0.14 ,\;\;\;\del n(1/2^+,1s1/2,n)=-0.12
$$
$$
\del n(3/2^+,0p3/2,p)=-0.56 ,\;\;\;\del n(1/2^+,0p1/2,p)=0.58
$$
     Therefore this state is predominantly a proton excitation.
     The first negative parity yrast levels have high excitation energy compared 
     with the corresponding experimental values. We obtained $E(1/2^-)=9.7 MeV$
     $E(3/2^-)=11.97 MeV$ (this state was obtained with $200$ Slater determinants
     and is not fully converged). The calculated $5/2^-$ is so high in energy that
     we cannot rule out a center of mass excitation. The variations of the 
occupation
     numbers compared with the ground state are
$$
\del n(1/2^-,0p3/2,n)=-0.16,\;\;\;\del n(1/2^-,0p1/2,n)=-0.7,
$$
$$
\del n(1/2^-,0d5/2,n)= 1.02,\;\;\;\del n(1/2^-,1s1/2,n)=-0.14
$$
$$
\del n(1/2^-,0p3/2,p)=-0.27,\;\;\;\del n(1/2^-,0p1/2,p)=0.33
$$
$$
\del n(3/2^-,1s1/2,n)=-0.84,\;\;\;\del n(3/2^-,1p3/2,n)=0.86
$$
     Therefore the $3/2^-$ state is a neutron excitation from the $sd$ shell to the $fp$ shell,
     while the $1/2^-$ is mostly an excitation from the $p$ shell to the $sd$ shell.  
\par
\subsection{ ${}^{16}C$,  ${}^{18}C$, ${}^{20}C$}
\par
     The experimental binding binding energy of ${}^{16}C$ is $110.75 MeV$ and the first
     excited state is $E(2^+)=1.766 MeV$. For this nucleus we used $300$ Slater determinants
     and obtained a binding energy of $114.707 MeV$ with a $0.6\%$ uncertainty. The theoretical
     $2^+$ has an excitation energy of $1.74 MeV$, in good agreement with the experimental value.
     This state is predominantly a neutron excitation since
$$
\del n(2^+,0d5/2,n)=0.12,\;\;\;\;\del n(2^+,1s1/2,n)=-0.12
\eqno(22)
$$
     The $fp$ shell is appreciably populated by $0.43$ neutrons and $0.33$ protons.
     It seems that intrashell excitations are overall in agreement with the experimental values
     (cf. the discussion of the other isotopes) but intershell excitations are too high
     compared with the experimental data.
\par 
     The experimental binding energy of ${}^{18}C$ is $115.67 MeV$ and $E(2^+)=1.59 MeV$.
     With $250$ Slater determinants we obtained a binding energy of $119.73 MeV$ and
     $E(2^+)=1.89 MeV$. The $2^+$ state is predominantly a neutron excitation with 
     a $0.12$ increase in the population of the $0d5/2$ orbital at the expenses of the
     $0d3/2$ and $1s1/2$. Also here we have $0.44$ neutrons and $0.3$ protons in the $fp$ shell.
\par
     The experimental binding energy of ${}^{20}C$ is $119.17 MeV$ and $E(2^+)=1.59 MeV$.
     With $200$ Slater determinants we obtained a binding energy of $124.43 MeV$ and
     $E(2^+)=1.94MeV$. The $2^+$ state is mostly a neutron excitation with 
     a $0.11$ decrease in the population of the $1s1/2$ orbital in favor  of the
     $0d3/2$ and $0d5/2/2$. The only appreciable change  in the number of proton is
     a $0.03$ decrease in the $0p3/2$ population in favor of the $0p1/2$ orbit. Also here we 
     have $0.46$ neutrons and $0.29$ protons in the $fp$ shell.
     We repeated this calculation using $400$ Slater determinants in order to see whether
     ${}^{22}C$ is more bound than ${}^{20}C$. Absolute values for the energies have
     a slower convergence with the number of Slater determinants than excitation energies,
     and $400$ Slater determinants are not sufficient to determine unambiguously whether
     ${}^{22}C$ is bound
     in this approach. We obtained $E_{gs}({}^{22}C)-E_{gs}({}^{20}C)= 0.2 MeV$ and this
     energy difference is  slowly decreasing with the number of Slater determinants.
     The model space used in this work can hardly properly describe halo nuclei.     
     The isotope ${}^{21}C$ is unbound by few MeV's.
\par
\subsection{ ${}^{17}C$,  ${}^{19}C$}
\par
     The experimental binding energy for ${}^{17}C$ is $111.48 MeV$. The ground state has
     $J^{\pi}=3/2^+$ and the known excited state have $E(1/2^+)=0.21 MeV$ and $E(5/2^+)=0.331MeV$ 
     For this nucleus we considered only $150$ Slater determinants and therefore the calculated
     binding energy is not well determined (we did not see in this case a linear behavior
     as a function of $1/N_{SD}$. The calculated binding energy is $114.48 MeV$. More importantly
     the ground-state has $J^{\pi}=1/2^+$ in disagreement with the experimental value.
     Although less accurate than the excitation energies in the previous cases, we have
     $E(3/2^+)=0.4 MeV$ and $E(5/2^+)=1.9 MeV$.
     The experimental binding energy for ${}^{19}C$ is $115.8 MeV$ and the ground-state  
     has $J^{\pi}=1/2^+$, the first excited state has $E(3/2^+)=0.196 MeV$ and the second
     excited state has $E(5/2^+)=0.269 MeV$. The calculated binding energy is $120.05 MeV$
     with an estimated uncertainty of $0.4\%$. The ground-state has $J^{\pi}=3/2^+$ in disagreement
     with the experimental value.
\par
\subsection{ Separation energies.}
\par
     Although we have seen, in this model space, a systematic underbinding for light isotopes and
     overbinding for the heavy ones, it is interesting to extract the neutron separation energies
     and to compare them with the experimental data. This is done in fig.6 . The overall trend
     is rather well reproduced, especially the even-odd effect. In all these calculations the   binding energies 
     are not fully converged,
\renewcommand{\baselinestretch}{1}
\begin{figure}
\centering
\includegraphics[width=10.0cm,height=10.0cm,angle=-90]{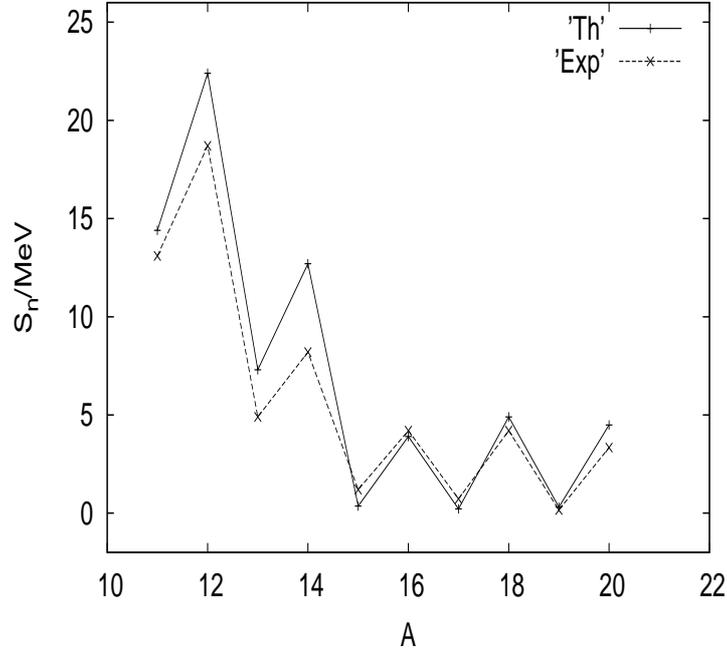}
\caption{Neutron separation energies for carbon isotopes.}
\end{figure}
\renewcommand{\baselinestretch}{2}
     that is, we need a larger number of Slater determinants. However this does not represent a problem
      as previously mentioned, since these calculations are variational. In other words, the theoretical
     errors have all the same sign and such errors tend to cancel out in the evaluation of the separation 
     energies. This seems to be especially true for the evaluation of the excitation energies.
     As a final point, let us mention that the sizes of the Hilbert spaces with $4$ major shells, range from 
     $3\times 10^{10}$
    (in the case of ${}^{10}C$) to about $10^{16}$ in the case of ${}^{22}C$
\par    
\section{ Summary.}
\par
     In this work we have studied carbon isotopes in a fully microscopic way using the chiral
     N3LO interaction properly renormalized to $4$ (in some cases $5$) major shells. In 
     this 
     treatment there are no adjustable parameters. We have evaluated binding energies, separation energies
     and few low energies levels. There seems to a systematic discrepancy with the experimental data
     whenever energy levels involve cross-shell excitation. Moreover, although by a small amount,
     ${}^{22}C$ is not bound. This is not very surprising since the model space is not well suited
     to describe loosely bound systems. The first $2^+$ state of 
     heavy even isotopes are dominated by neutron excitation
     and for light odd isotopes the proper spin of the  ground state  is reproduced. 

\vfill
\eject
\end{document}